\documentclass[twocolumn,showpacs,aps,floats]{revtex4}
\usepackage{graphicx}%
\usepackage{dcolumn}
\usepackage{latexsym}
\usepackage{amssymb}
\usepackage{amstext}
\usepackage{enumerate}
\usepackage{verbatim}
\usepackage{rotating}
\usepackage{epsfig}

\begin{document}
\title{Synchronization and Coarsening (without SOC) in a Forest-Fire Model}
\author{K. E. Chan, P. L. Krapivsky, and S. Redner}
\affiliation{Center for BioDynamics, Center for Polymer Studies, 
and Department of Physics, Boston University, Boston, MA 02215}


\begin{abstract}   
  
  We study the long-time dynamics of a forest-fire model with deterministic
  tree growth and instantaneous burning of entire forests by stochastic
  lightning strikes.  Asymptotically the system organizes into a coarsening
  self-similar mosaic of synchronized patches within which trees regrow and
  burn simultaneously.  We show that the average patch length $\langle
  L\rangle$ grows linearly with time as $t\to\infty$.  The number
  density of patches of length $L$, $N(L,t)$, scales as $\langle
  L\rangle^{-2} {\cal N} (L/\langle L\rangle)$, and within a mean-field rate
  equation description we find that this scaling function decays as ${\cal
    N}(x)\sim e^{-1/x}$ for $x\to 0$, and as $e^{-x}$ for $x\to\infty$.  In
  one dimension, we develop an event-driven cluster algorithm to study the
  asymptotic behavior of large systems.  Our numerical results are consistent
  with mean-field predictions for patch coarsening.

\end{abstract}
\maketitle


\section{Introduction}
\label{sec:intro}

Forest-fire models \cite{gh,bak90,Henley,dro92} are simple archetypical
examples of driven dissipative systems that exhibit intriguingly rich
spatio-temporal structures \cite{dro92,all93,gra93,cla97,hon97,new02}.  These
models provide a simple paradigm for cooperative time-dependent phenomena,
such as epidemics, oscillatory chemical reactions, electrical neuron
activity, cardiac dynamics, and turbulence \cite{joh,excite,mer,bak90,bak01}.
As the name indicates, the forest-fire model ostensibly describes the time
evolution of burning trees in a forest.  In typical models of this genre,
trees are located at regular lattice sites and each can exist in one of three
states: burnt, alive, or burning.  The dynamics involves the following
elements: (i) A burnt tree turns into a living tree at some specified rate;
(ii) a living tree can be ignited, either by a lightning strike or by fire
spreading from a neighboring burning tree; (iii) after a specified time
interval a burning tree is consumed and the fire at this location is
extinguished.

Depending on which of these processes are operative, as well as their
relative rates, different dynamical behaviors can arise, ranging from
self-organized critical behavior with fires of all sizes occurring
\cite{bak90,Henley,dro92}, to spiral fire-front wave propagation
\cite{bak90,GK91}.  While forest fire models have been extensively
investigated, there is still uncertainty about their long-time properties
even after extensive numerical simulations in many realizations of the model
\cite{new02}.

This work is focused on a specific version of the forest-fire model which
exhibits coarsening \cite{bray} rather than self-organized criticality or
complex fire front propagation.  Because of this phenomenological simplicity,
we can apply the rate equations in a natural way to determine the evolution
of the system.  The model itself was first introduced by Drossel
\cite{dro96}.  Its crucial feature is that tree growth is deterministic; a
tree that has just burned remains dead for exactly one time unit and then a
new tree reappears.  This particular regrowth rule is the mechanism that
gives rise to a coarsening mosaic of growing synchronized forests.  This is a
generic feature and does not require the tuning of model parameters to
critical values.

Let us define a ``patch'' as a coherent region of the system that is either
occupied by live trees or by burnt trees.  This patch evolves by tree
regrowth and by the burning of trees due to lightning strikes.  The
mechanism for coarsening is that adjacent patches must eventually
synchronize, after which they evolve in phase \cite{dro96}.  While
neighboring patches begin their existence as distinct, eventually the burnt
trees in one patch will regenerate while the adjacent patch is still
forested.  When this occurs, all the trees in these two patches become
incorporated into an augmented patch which then evolves as a single unit.
(Fig.~\ref{fig:model}).

In Ref.~\cite{dro96}, the process was investigated numerically and the total
number of patches $N(t)$ was found to decay with time.  However, more
quantitative observations were not reported.  We will show that $N(t)\propto
t^{-1}$ and we will investigate the patch length distribution, both
analytically and numerically.  For our analytic study, we will employ the
classical rate equation of aggregation kinetics \cite{ernst}.  This approach
is ideally suited to treat the coarsening behavior of the system under
investigation.

Another important feature of our approach is that we treat the dynamics at a
mesoscopic level in which the basic units are synchronized patches rather
than individual trees.  In the long-time limit, we will argue that the
lifetime of such patches can be viewed as deterministic.  By tracking only
the merging events of adjacent patches we are able to investigate systems of
effectively much larger size and to much longer times than that accessible by
tree-based simulations.  In addition to obtaining a power-law growth of the
average patch length, our method yields clean results for the probability
distribution of patch lengths.  This distribution is found to obey scaling,
with no memory of the initial state retained and with the asymptotes of the
distribution in good agreement with rate equation predictions.  While we
focus on the particular case of one dimension, our approach should also apply
in higher dimensions.

In the next section, we define the model and outline the effective mesoscopic
picture for the evolution of patches.  In Sec.\ III, a rate equation
description for this evolution is presented and basic results about the patch
length distribution are obtained.  We explain our simulational approach and
describe the results that follow in Sec.\ IV.  Our basic conclusions are
given in the last section.

\section{Forest Evolution in One Dimension}
\label{sec:1d}

The evolution of the system is governed by the competition between two
fundamental time scales.  Suppose that each tree in the system may be struck
by lightning at a rate $\Lambda$.  Then a forest of length $L$ has a
characteristic lifetime $t_{\rm occ}\sim (\Lambda L)^{-1}$ before one of its
trees is struck by lightning.  We assume that the time to burn the forest
completely is much less than any other time scale in the problem, so that we
view the burning of a forest as instantaneous.  There is also the
deterministic time interval $t_{\rm emp}$ between the instant that a forest
burns down and the reappearance of trees in this burnt patch.  We assume that
this refractory time is the same for all trees, so that regrowth of trees in
a single burnt patch occurs simultaneously.  Without loss of generality, we
take this refractory time to be $t_{\rm emp}=1$.

\begin{figure}[ht]
 \epsfig{file=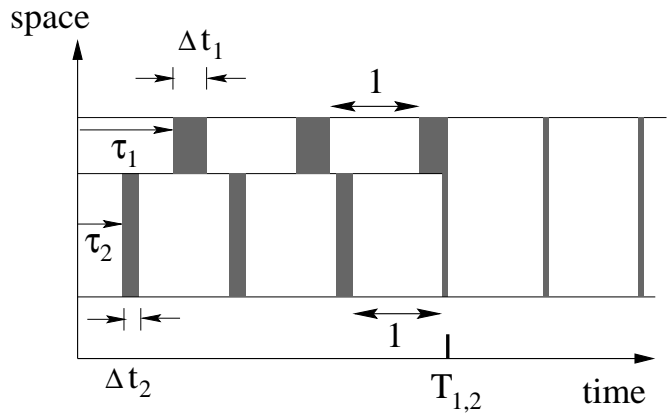, width=8.75truecm}
\caption{Illustration of the  merging of two adjacent patches of lengths $L_1$
  and $L_2>L_1$.  White space indicates a burned patch while the shaded
  region indicates a forested patch.  The times until the first regrowth
  events are $\tau_1$ and $\tau_2$.  Each forested patch survives for a time
  $\Delta t_j\propto L_j^{-1}$ until lightning strikes and instantaneous
  burning occurs.  The patch then regrows after exactly one time unit
  elapses.  Because $\Delta t_1$ and $\Delta t_2$ are different, the two
  patches will eventually synchronize at the joining time $T_{1,2}$.  }
\label{fig:model}
\end{figure}

At early times, where $t_{\rm emp}\leq t_{\rm occ}$, the stochastic nature of
lightning events is important.  However, as we shall soon show, patches
naturally grow with time.  Thus $t_{\rm emp}$ eventually becomes much larger
than $t_{\rm occ}$ and the fire dynamics becomes nearly deterministic in the
long-time limit.  To describe this late-stage dynamics, we ignore individual
trees and treat the system mesoscopically as a contiguous array of patches,
each with length $L_j$.  Each patch can either be forested or burnt.  If two
distinct but adjoining forested patches arise by the regrowth of one patch
next to a forested patch, they immediately join to form a larger patch
(Fig.~\ref{fig:model}).

At long times, patches have a small lifetime and they are almost always in
the burnt state.  Without loss of generality we initialize the system so that
it is effectively in this long-time state.  That is, at $t=0$, we consider
all patches to be burnt, and we define $\tau_j$ (with $0<\tau_j<1$) to be the
time at which the $j^{\rm th}$ burnt patch first becomes a forest.  Consider
now two adjacent patches, and let $\phi_j=\tau_{j+1}-\tau_j$ be the time
difference between the appearance of forest $j$ and forest $j+1$.  After
these two forests undergo one cycle of regrowth and subsequent burning,
$\phi_j$ changes by $\Delta t_{j+1}-\Delta t_j$, where $\Delta t_j=(\Lambda
L_j)^{-1}$ is the lifetime of the $j^{\rm th}$ forest.  This shift in the
difference of burning times continues until $\phi_j$ reaches either
$\phi_j=0$ or $\phi_j=1$.  When this occurs, the forests necessarily join and
are subsequently synchronized (Fig.~\ref{fig:model}).

Since lightning strikes a forest of length $L_j$ at rate $\Lambda L_j$, the
lifetime $\Delta t_j$ of each forest is a stochastic variable whose average
value is $\langle\Delta t_j\rangle=(\Lambda L_j)^{-1}$.  In the long time
limit, these lifetimes become very small.  Therefore we may replace the sum
of a large number of these lifetimes by the average lifetime times the number
of cycles.  Thus for two adjacent patches $j$ and $j+1$ that satisfy
$L_j<L_{j+1}$ and $\tau_j<\tau_{j+1}$, the joining time $T_{j,j+1}$ is
\begin{eqnarray}
\label{eq:JT}
T_{j,j+1}  =\left\{\begin{array}{ll}
{\displaystyle \frac{\tau_{j+1}-\tau_j}
                 {\langle\Delta t_{j+1}\rangle - \langle\Delta t_j\rangle}},
          &\quad \tau_j<\tau_{j+1}, \\
{\displaystyle \frac{1-\tau_j+\tau_{j+1}}
                 {\langle\Delta t_j\rangle - \langle\Delta t_{j+1}\rangle}},
          &\quad \tau_j>\tau_{j+1}.
       \end{array}\right.
\end{eqnarray}

An analogous result holds when $L_j>L_{j+1}$.  From these joining times
between adjacent patches, we conclude that the typical joining time scales
with the average patch length as $T\sim \langle\Delta t\rangle^{-1}\sim
\Lambda\langle L\rangle$.  Thus in a time interval $dt\sim T$, a typical
patch grows by an amount $dl\sim L$.  Consequently ${d\over dt}\langle L\rangle\sim
T^{-1}\langle L\rangle$, and we obtain
\begin{equation}
\label{ellt}
\langle L\rangle \sim \Lambda^{-1}t.
\end{equation}

In previous forest-fire models that are driven by stochastic tree growth and
by stochastic lightning strikes \cite{Henley,dro92}, the latter rate needs to
be very small to ensure non-trivial dynamics.  There is no need for such
parameter tuning in the present model, as the magnitude of $\Lambda$
determines only the overall scaling of the typical forest length.  Therefore,
we set $\Lambda=1$ henceforth.

\section{Rate Equation Description}
\label{sec:RE}

A natural approach to determine the evolution of the patch length
distribution is the rate equations.  Let $N(L,t)$ be the number density of
patches of length $L$ at time $t$, and let $N(t)=\int _0^\infty N(L,t)\,dL$
be the total number of patches of any length.  In the rate equation
description of these quantities, we make the mean-field assumption that there
are no correlations between adjacent patches.  In a similar spirit, we also
ignore the initial phase difference between two patches in
Eqs.~(\ref{eq:JT}), so that the joining rate of two patches is simply
proportional to $\langle\Delta t_j\rangle - \langle\Delta
t_{j+1}\rangle\propto |L_j^{-1}-L_{j+1}^{-1}|$.  With these approximations,
the patch length distribution evolves according to the rate equation
\begin{eqnarray}
\label{smol} 
{\partial N(L,t)\over \partial t}
&=&{1\over 2 N(t)}\int_0^L dl\,K(l,L-l)\,N(l,t)\,N(L-l,t)\nonumber\\
&-&{N(L,t)\over N(t)}\int_0^\infty dl\,K(L,l)\,N(l,t).
\end{eqnarray}
Here $K(x,y)=|x^{-1}-y^{-1}|$ is the rate at which a patch of length $x$
joins with a patch of length $y$.  

This rate equation is nearly identical in form to the corresponding equation
for the kinetics of aggregation \cite{ernst}, except for the overall factor
of $1/N(t)$.  This difference arises because we consider a finite system and
we track the number of forests of a given length rather than the corresponding
probability.  However, this factor can be absorbed into a rescaled time
variable defined by
\begin{equation}
\label{newt} 
{\cal T}=\int_0^t {dt'\over N(t')},
\end{equation}
to reduce Eq.~(\ref{smol}) to the standard form of the rate equation for
aggregation.  This can then be analyzed by well-established methods
\cite{ernst}.

While the rate equation with reaction rate $K(x,y)=|x^{-1}-y^{-1}|$ has not
been solved, basic features about the long-time solution can be inferred from
a scaling approach.  In general, the long-time behavior of the rate equations
with homogeneous reaction rates that satisfy: (i) $K(ax,ay)=a^\lambda
K(x,y)$, and (ii) $K(x,y)\sim x^\mu y^\nu$ for $x\ll y$ ($\lambda=\mu+\nu$),
have been generically classified according to whether $\mu>0$, $\mu<0$, or
$\mu=0$ \cite{ernst,van}.  In all cases, the asymptotic length distribution
approaches a scaling form
\begin{equation}
\label{scaling}
N(L,t)\simeq \langle L(t)\rangle^{-2}{\cal N}(L/\langle L(t)\rangle),  
\end{equation}
in which the average patch length grows algebraically with time, $\langle
L\rangle\sim {\cal T}^{1/(1-\lambda)}$, when $\lambda<1$.  However, the
scaling function ${\cal N}$ exhibits different behaviors in the three cases.

The reaction rate for our forest-fire model, $K(x,y)=|x^{-1}-y^{-1}|$, is
homogeneous with homogeneity exponent $\lambda=-1$, while $\mu=-1$ and thus
$\nu=0$.  Therefore, the scaling theory prediction for the average patch
length becomes $\langle L\rangle\sim {\cal T}^{1/2}$.  Consequently, $N({\cal
  T})\sim \langle L\rangle^{-1}\sim {\cal T}^{-1/2}$, and from
Eq.~(\ref{newt}), we recover $\langle L\rangle\sim t$, in agreement with the
qualitative argument preceding Eq.~(\ref{ellt}).  According to the general
classification scheme of van Dongen and Ernst \cite{ernst,van}, since
$\mu=-1$, the scaling function ${\cal N}(x)$ should vanish exponentially in
the limits of small and large $x$:
\begin{eqnarray}
\label{limits}
{\cal N}(x) \sim \left\{ \begin{array}{ll}
             e^{-1/x}, & \quad x\to 0, \\
             e^{-x}, & \quad x\to\infty. 
           \end{array}\right.
\end{eqnarray}

For short patches, this result therefore predicts $N(L,t)\sim e^{-t/L}$.
This behavior can also be established directly from Eq.~(\ref{smol}).  When
$L\ll t$, the gain term in the rate equation can generally be ignored.
Additionally, in this limit the reaction rate $K(L,l)$ simplifies to
$L^{-1}$.  Hence the density of short patches satisfies ${\partial N\over
  \partial t}=-L^{-1}N$, which indeed implies the above exponential decay.

\section{Simulation Results}
\label{sec:sim}

In our simulations, we start the system with a random array of patches of
lengths $\{L_j(t=0)\}$.  As discussed in Sec.~II, it is asymptotically exact
to replace the stochastic forest lifetime $\Delta t_j$ by its average value
$\langle\Delta t_j\rangle=(L_j)^{-1}$.  The time for regrowth of a tree is
always equal to one.  Thus the dynamical steps become deterministic and
randomness enters only through the initial conditions.

For convenience, we assume that each patch is initially in the burned state.
{}From these regrowth and burning processes, a simulation of forest fires
should be based on the following steps:

\begin{enumerate}
  
\item Initialize the line with patches of random lengths $L_j$ and with
  random times $\tau_j$, for $j=1, 2, \ldots, N$ at which the $j^{\rm th}$
  patch first turns into a forest.  Assign a lifetime $(L_j)^{-1}$ to a
  forest of length $L_j$.
  
\item Use Eq.~(\ref{eq:JT}), compute the joining times $T_{j,j+1}$ for all
  neighboring pairs of patches.
  
\item Sort the list of joining times $\{T_{j,j+1}\}$ in ascending order.  A
  standard sort algorithm \cite{recipes} requires a time of the order of
  $N\ln N$ for a set of $N$ elements.
  
\item Join the pair of patches $(j,j+1)$ with the minimal joining time and
  increment the time accordingly.  Recompute the joining times $T_{j-1,j}$
  and $T_{j,j+2}$ of the patches adjacent to the newly-joined forest.
  
\item Decrement the total number of patches by 1 and return to step 3.

\end{enumerate}

Such an algorithm is perforce inefficient because of the re-sorting of
joining times after each event.  However, this step is typically unnecessary
for two reasons.  First we ``cache'' only a small subset of the joining times
with $T_{j,j+1}$ less than a judiciously-chosen cutoff time $T_c$ and sort
only this subset in step 3.  We need not consider joining times $T_{j,j+1} >
T_c$ because these events in the far future will never be considered before
the current list of joining times needs to be re-sorted.  This restriction
significantly reduces the time needed to sort the joining time list for large
$N$.

Second, it is unnecessary to re-sort this reduced joining time list after
each joining event because only the two joining times $T_{j-1,j}$ and
$T_{j,j+2}$ are modified.  If these updated joining times are greater than
the elements in the already-sorted list, there is no need for re-sorting.
One can simply continue to use the joining times from the pre-sorted list to
define joining events until one of the newly-created joining times becomes
less than the next joining time in the pre-sorted list.  Only when such a
mis-ordering occurs is it necessary to return to step 3 and re-sort the
joining time list.  These steps are completely analogous to those employed in
Ref.~\cite{KRL95} to simulate the kinetics of one-dimensional ballistic
annihilation reactions efficiently.

\begin{figure}[ht]
  \epsfig{file=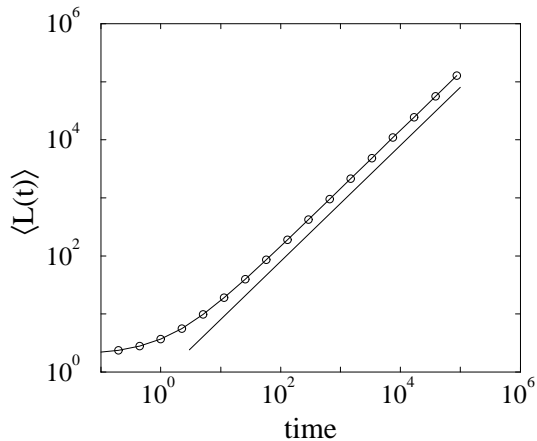, width=7truecm}
 \caption{Average patch length $\langle L(t)\rangle$ versus
   time $t$ for initial length distribution $N(L,t=0)=L_0^{-1}
   \exp{(-L/L_0)}$, with $L_0=0.1$.  A straight line of slope 1 is shown for
   comparison.   }
 \label{fig:Lav}
\end{figure}

\begin{figure}[ht]
  \epsfig{file=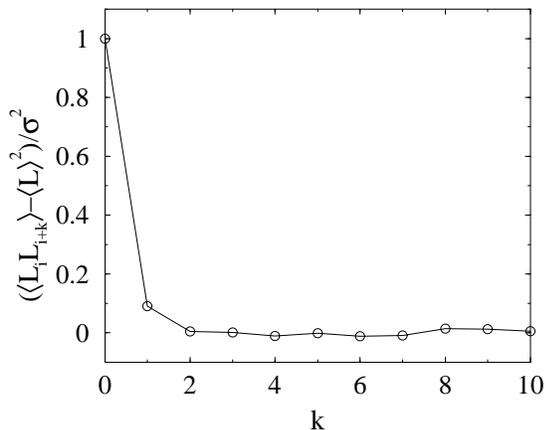, width=7truecm}
 \caption{Dependence of the normalized correlation function 
   $(\langle L_i L_{i+k}\rangle-\langle L\rangle^2)/\sigma^2$, where
   $\sigma^2=\langle L^2\rangle-\langle L\rangle^2$, on patch separation $k$
   at $t=2.25^9\approx 1478$.  }
 \label{fig:corr}
\end{figure}

\begin{figure}[ht]
  \epsfig{file=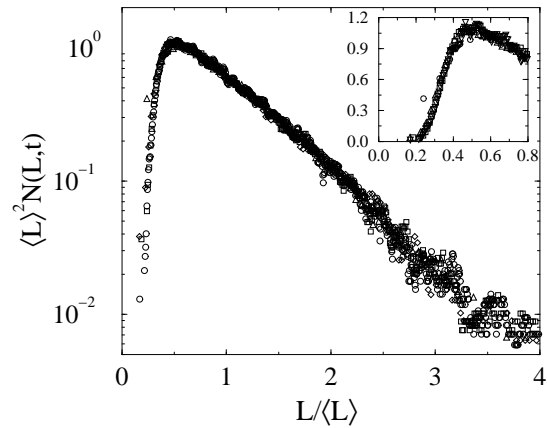, width=7truecm}
 \caption{Number of patches  $N(L,t)$ of length $L$ at $t=1478$, 3325, 7482,
   and 16384 ($t=2.25^n$ with $n=9-12$) plotted in scaled form.  The main
   plot shows the data on a semi-logarithmic scale to illustrate the
   exponential decay of the large-length tail.  The inset shows the same data
   in the small-length limit on a linear scale to highlight the essentially
   singular form.  }
 \label{fig:N-scaled}
\end{figure}

We initialize the system with $N = 5\times10^6$ patches whose lengths are
randomly drawn from the distribution $N(L,t=0)= L_0^{-1} \exp{(-L/L_0)}$,
with $L_0=0.1$.  This initial length should be viewed as much larger than the
lattice spacing between individual trees.  We are interested in the
intermediate asymptotic regime, where the average patch length is growing
systematically with time and before finite-size effects begin to play a role.
Fig.~\ref{fig:Lav} indicates that this intermediate asymptotic regime begins
when $t >t^*\approx 10$.  An important feature of the system at long times is
that there is only a very short-range spatial correlation in the lengths of
neighboring patches (Fig.~\ref{fig:corr}).  In particular, the normalized
correlation function $(\langle L_i L_{i+k}\rangle - \langle
L\rangle^2)/\sigma^2$, with $\sigma^2=\langle L^2\rangle-\langle L\rangle^2$,
quickly approaches zero for $k\geq 2$.  This provides empirical justification
for the validity of the mean-field approximation of the rate equations.  It
is worth remarking that such lack of spatial correlations appear ubiquitously
in many one-dimensional coarsening processes \cite{bray,CB}.

We now examine the behavior of $N(L,t)$ for representative values of $t>t^*$
(Fig.~\ref{fig:N-scaled}).  The different sets exhibit data collapse
according to the scaling ansatz of Eq.~(\ref{scaling}).  The large-length
tail of this distribution appears to be consistent with a simple exponential
decay.  However, there is a very small but apparently systematic downward
curvature to the data for which we do not have an explanation.  The
small-length tail decays extremely rapidly near the origin and there are
essentially no patches with scaled length less than 0.2.  This is consistent
with the essential singularity predicted by Eq.~(\ref{limits}).  Again, it is
worth remarking that in almost all coarsening processes that are controlled
by an underlying diffusion process, the small-length tail of the patch-length
distribution is a linear function.  The one well-known example of an
essential singularity in the small-size tail of a cluster distribution is the
aggregation of Brownian particles \cite{ernst}.

\section{Conclusions}

We have developed a mesoscopic description for a forest-fire model with
stochastic lightning strikes and deterministic tree growth.  Instead of
treating the system at the level of single trees, the basic element in our
description are patches of synchronized trees.  Each patch undergoes periodic
cycles of burning and regrowth, and in the long-time limit, the lifetime of
patches can be viewed as deterministic.  Whenever two adjoining patches are
simultaneously in the forested state, the patches join and remain
synchronized forever.  The number of distinct patches decreases while their
typical length grows continuously with time as in classical coarsening
processes \cite{bray}.

This mesoscopic description is well-suited to a rate equation approach for
the evolution of the patch-length distribution, as well as efficient
simulations.  Numerically, we find that the average patch length grows
linearly with time, while the number of patches correspondingly decreases as
$1/t$.  The patch length distribution is sharply peaked, with an exponential
large-length tail and an essential singularity in the small-length tail.
These features are consistent with the rate equation predictions.  Even
though the system has nearest-neighbor interactions only, there are
essentially no spatial correlations in the lengths of neighboring patches.
Because of this lack of spatial correlation, we can anticipate that the rate
equation predictions, which are based on no correlations between patch
lengths, should provide an accurate account of the one-dimensional
simulations.

Most aspects of our approach can be extended to higher spatial dimensions
$d$.  A complicating factor in developing numerical simulations is that the
number of neighbors for a given patch is variable.  Nevertheless, the same
updating rule given by Eqs.~(\ref{eq:JT}) will still apply, with patch length
being replaced by patch volume.  As a result, we expect that the average
patch volume should grow linearly with time.  Under the assumption that
patches remain compact, this would imply that the typical length scale of a
growing patch would grow in time as $L\sim t^{1/d}$.  It is of course far
from obvious that patches remain compact.  Understanding of the forest-fire
model in higher dimensions appears to be an interesting challenge.

Finally, we want to stress that the seemingly innocent change of the tree
growth rule from stochastic to deterministic drastically affects the
dynamics.  Almost all earlier work focused on stochastic tree growth.
According to the literature on such model, self-organized critical behavior
should occur in the limit of infinitesimal rate of lightning.  However, after
considerable effort, the understanding of the scaling laws is still
incomplete, and even their very existence has recently been questioned
\cite{new02}.  In contrast, when the tree growth is deterministic, there is
no need to tune the rate of lightning strikes to zero, and the model exhibits
simple coarsening rather than complex time-dependent phenomena.  It is of
course possible that the higher-dimensional version of the model will offer
some surprises.

\medskip

We are grateful to NSF grant DMR9978902 for partial support of this work.


\begin{thebibliography}{99}
  
\bibitem{gh}     J.~M.~Greenberg and S.~P.~Hastings, 
                 SIAM J.\ Appl.\ Math. {\bf 34}, 515 (1978).
\bibitem{bak90}  P.~Bak, K.~Chen, and C.~Tang,  
                 Phys.\ Lett.\ A {\bf 147}, 297 (1990);
                 K.~Chen, P.~Bak, and M.~Jensen,
                 Phys.\ Lett.\ A {\bf 149}, 207 (1990).
\bibitem{Henley} C.~L.~Henley, Bull.\ Am.\ Phys.\ Soc. {\bf 34}, 838 (1989); 
                 Phys.\ Rev.\ Lett.\ {\bf 71}, 2741 (1993).
\bibitem{dro92}  B.~Drossel and F.~Schwabl,  
                 Phys.\ Rev.\ Lett.\ {\bf 69}, 1629 (1992); for a 
                 review, see S.~Clar, B.~Drossel, and F.~Schwabl, 
                 J. Phys.: Condens.\ Matter {\bf 8}, 6803 (1996).
\bibitem{all93}  K.~Christensen, H.~Flyvbjerg, and Z.~Olami, 
                 Phys.\ Rev.\ Lett.\ {\bf 71}, 2737 (1993).
\bibitem{gra93}  P.~Grassberger, J. Phys.\ A {\bf 26}, 2081 (1993).
\bibitem{cla97}  S.~Clar, K.~Schenk, and F.~Schwabl, 
                 Phys.\ Rev.\ E {\bf 55}, 2174 (1997);
                 S.~Clar, B.~Drossel, K.~Schenk, and F.~Schwabl, 
                 Phys.\ Rev.\ E {\bf 56}, 2467 (1997).
\bibitem{hon97}  A.~Honecker and I.~Peschel, Physica A {\bf 239}, 509 (1997). 
\bibitem{new02}  K.~Schenk, B.~Drossel, and F.~Schwabl, 
                 {\it cond-mat}/0105121;
                 G.~Pruessner and H.~J.~Jensen, {\it cond-mat}/0201306;
                 P.~Grassberger, {\it cond-mat}/0202022.
\bibitem{joh}    A.~Johansen, Physica D {\bf 78}, 186 (1994);
                 J.\ Theor. Biol. {\bf 178}, 45 (1996).
\bibitem{excite} J. J. Tyson and J. Keener, Physica D {\bf 32}, 327 (1988).
\bibitem{mer}    E. Meron, Phys.\ Rep.\ {\bf 218}, 1 (1992).
\bibitem{bak01}  P.~Bak, K.~Chen, and M.~Paczuski,  
                 Phys.\ Rev.\ Lett.\ {\bf 86}, 2475 (2001).
\bibitem{GK91}   P. Grassberger and H. Kantz, 
                 J. Stat.\ Phys.\ {63}, 685 (1991).
\bibitem{bray}   For a general review of coarsening, see e.g.  
                 see A.~J.~Bray, Adv.\ Phys. {\bf 43}, 357 (1994).
\bibitem{dro96}  B.~Drossel, Phys.\ Rev.\ Lett.\ {\bf 76}, 936 (1996).
\bibitem{ernst}  For a general review of the rate equations of aggregation, 
                 see M.~H.~Ernst, in {\it Fractals in Physics}, 
                 edited by L.~Pietronero and E.~Tosatti 
                 (Elsevier, Amsterdam, 1986), p.~289.
\bibitem{van}    P.~G.~J.~van~Dongen and M.~H.~Ernst, 
                 Phys.\ Rev.\ Lett.\ {\bf 54}, 1396 (1985).
\bibitem{recipes} W.~H.~Press, B.~P.~Flannery, S.~A.~Teukolsky, 
                 and W.~T.~Vetterling, {\it Numerical Recipes}
                 (Cambridge University Press, Cambridge, 1986).
\bibitem{KRL95}  P.~L.~Krapivsky, S.~Redner, and F.~Leyvraz, 
                 Phys.\ Rev.\ E {\bf 51}, 3977 (1995).
\bibitem{CB}     S. J. Cornell and A. J. Bray, Phys.\ Rev.\ E {\bf 54}, 1153
                 (1996).


\end{thebibliography}
\end{document}